\documentclass[oribibl]{llncs}

\PassOptionsToPackage{hyphens}{url}
\usepackage[colorlinks = true,
  linkcolor = blue,
  citecolor = blue,
  urlcolor = blue]{hyperref}

\usepackage{url}
\usepackage{csquotes}
\usepackage{multirow}
\usepackage{amsmath}
\usepackage{amssymb}
\usepackage{graphicx}
\usepackage{tabularx}
\usepackage{booktabs}
\usepackage{threeparttable}

\begin{document}

\title{A Systematic Literature \\ Review on the NIS2 Directive}

\author{Jukka Ruohonen\orcidID{0000-0001-5147-3084} \\ \email{juk@mmmi.sdu.dk}} \institute{University of Southern Denmark, S\o{}nderborg, Denmark}

\maketitle

\begin{abstract}
The second network and information security (NIS2) directive was enacted in the
European Union (EU) in late 2022. It deals particularly with European critical
infrastructures, enlarging their scope substantially from an older directive
that only considered the energy and transport sectors as critical. The
directive's focus is on cyber security of critical infrastructures, although
together with other new EU laws it expands to other security domains as
well. Given the importance of the directive and most of all the importance of
critical infrastructures, the paper presents a systematic literature review on
academic research addressing the NIS2 directive either explicitly or
implicitly. According to the review, existing research has often framed and
discussed the directive with the EU's other cyber security laws. In addition,
existing research has often operated in numerous contextual areas, including
industrial control systems, telecommunications, the energy and water sectors,
and infrastructures for information sharing and situational awareness. Despite
the large scope of existing research, the review reveals noteworthy research
gaps and worthwhile topics to examine in further research.
\end{abstract}


\section{Introduction}

Europe, like all regions in the world, is facing an ever-increasing trend of
digitalization. With increasing digitalization comes also increasing cyber
security risks. At the same time, geopolitical conflicts and tensions have
emerged throughout the world, and technological development progresses
rapidly. These and other ``mega-trends'' affect cyber security too. Given this
state of affairs, the EU has recently adopted many new cyber security
regulations. Among these is the NIS2 directive. It repeals an older 2016
directive that addressed the cyber security of network and information systems
(NIS). The older directive was an important historical part in the evolution of
the EU's overall cyber security framework. In terms of public sector
administration and governance, it was built upon national computer security
incident response teams (CSIRTs), which too are an important part of the global
evolution of cyber security governance~\cite{Skopik16}. The new NIS2 directive
builds upon this evolutionary trajectory, further strengthening CSIRTs and their
coordination with private sector actors.

Coordination is important because many---if not the most---of the Europe's
critical infrastructures are either owned or operated by private sector
entities. The NIS2 directive categorizes these entities into two groups:
essential and important. More requirements are imposed upon the former group. In
general, the demarcation between the two groups is done by relying on the
concept of criticality, which in the NIS2's Article~2 mostly refers to
``critical societal or economic activities''. With this concept, eleven sectors
of high criticality are listed in the directive's Annex I. These include
traditional critical infrastructure sectors, such as the energy and transport
sectors, as well as banking, healthcare, and central government
administration. In addition, Annex~II provides a further listing of other
critical sectors, including postal services, waste management, research
institutions, and providers of digital services, whether online marketplaces or
search engines. Already these exhaustive listings make the NIS2 directive highly
relevant for the research domain investigating critical infrastructures.

Many reviews (see, e.g., \cite{Mikac23}) have already been published about the
NIS2 directive and its relation to critical infrastructure protection in
Europe. However, thus far, a review has been lacking about the research
addressing NIS2 either explicitly or implicitly. Against this backdrop, the
paper presents a systematic literature review~(SLR) about NIS2-related
research. The goal is thus not a comprehensive review of the NIS2 directive
itself---instead, the motivation and focus are on the aspects existing research
have considered relevant or interesting about the directive, and on which
particular contexts the directive has been discussed in the academic
literature. On these notes, Section~\ref{sec: methods} elaborates the SLR
approach used for retrieving the literature. Then, the review is presented in
Section~\ref{sec: review} after which the conclusion follows in
Section~\ref{sec: conclusion}.

\section{Methods}\label{sec: methods}

The search procedure for the SLR was simple: the databases of six scientific
publishers were queried with a string ``NIS2 AND directive'', where AND is a
Boolean operator. The word directive was necessary because the acronym NIS2
refers to other things in biomedical and related sciences. With two exceptions,
the searches were restricted to abstracts. The first exception is the Elsevier's
ScienceDirect database for which a search was restricted to abstracts, titles,
and keywords supplied by authors. The second exception is the Springer's
database that does not allow searches to be restricted to specific elements in
publications. For this database, ``NIS2 directive'' (with the quotation marks)
was used as a search term. With respect to SLRs in general, these two exceptions
demonstrate that the long-lasting problem with publishers' databases is still
present~\cite{Kitchenham09}. Then, regarding the other crucial factor in~SLRs,
the exclusion criteria, publications were only qualified insofar as they
(1)~were peer reviewed, (2) were written in English, and (3)~discussed the NIS2
directive explicitly with more than three sentences. The last exclusion
criterion ensured that publications merely mentioning the NIS2 directive in
passing did not enter into the sample.

\begin{figure}[th!b]
\centering
\includegraphics[width=\linewidth, height=6cm]{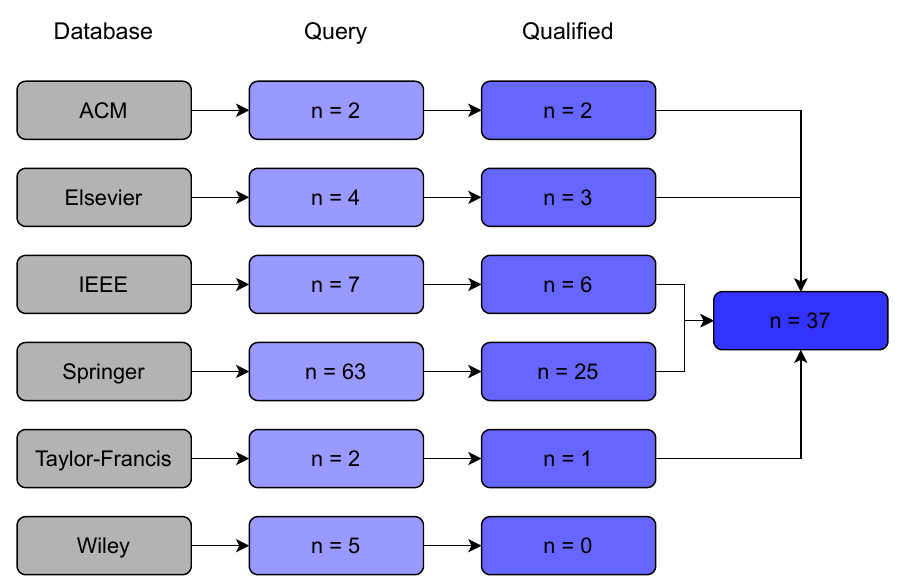}
\caption{The Literature Search}
\label{fig: sample}
\end{figure}

As can be seen from the illustration in Fig.~\ref{fig: sample}, the final sample
contains $37$ publications. Most of these were published by
Springer. Conventional scientific articles and publications in peer reviewed
conference proceedings are the most common publication types, although the
sample contains also a few monographs and edited collections. The reason for the
relatively small amount of papers is likely partially explained by the fact that
NIS2 is still rather new. This point manifests itself also in that all
publications sampled are relatively new; the majority was published in
2024. Although the deadline for national transpositions of the NIS2 directive
was in October 2024, as specified in the NIS2's Article~41, the implementation
work continues also thereafter. For instance, the deadline for the member states
to identify their critical entities is in April 2025 according to
Article~3(3). Already these deadlines and other schedules imply that practical
evaluation research is missing from the sample. While critical reflections have
been published, it is understood in the literature that evaluations can only be
done after the directive has been operational for a sufficient amount of
time~\cite{springer2}. Against this backdrop, it seems fair to assume that in
the future also evaluation work will be published, whether the work is about
legal analyses of national adaptations, administration and enforcement, or the
actual cyber security impact. In any case, the little below forty papers avoids
the common feasibility~\cite{Kitchenham09}, or breadth and depth \cite{Fisch18},
obstacles often encountered in SLRs.

\section{Review}\label{sec: review}

The review of the $37$ publications is presented by focusing on two broad
themes: the framing presented in the publications with respect to other EU laws
and the EU's jurisprudence in general, and the technological or other
application contexts the publications have focused on. The former theme is
important because it helps to understand how the publications have generally
understood and approached the EU's cyber security governance framework. It is
also important because the sample contains legal scholarship too. Then, the
second theme is relevant because it allows to understand the contexts academic
research has considered relevant or interesting in terms of the NIS2
directive. Both themes are also relevant regarding the common rationale behind
SLRs, the examination whether there are notable gaps in
existing~knowledge. Before continuing, a remark about presentation should be
made: because the review mixes also a few relevant publications that are not
part of the actual sample, the publications belonging to the SLR \textit{per~se}
are enumerated in two tables soon presented.

\subsection{Legal Framings}

The legal framings done in the publications are summarized in Table~\ref{tab:
  legal framings}. As can be seen, quite a few other EU laws have been
frequently discussed in conjunction with the NIS2 directive. The large amount of
related laws discussed prompts the first critical reflection: many publications
have merely pointed out or even enumerated EU laws without any deeper
analysis. While it is important to acknowledge the existence of regulations, it
remains debatable what scholarly purposes brief listings or footnote-like
mentions serve. The second point is that some publications have connected the
large amount of EU laws to an argument that complexity and regulatory
fragmentation have increased~\cite{taylorfrancis1, springer16, springer2}. The
third point is that only five publications have discussed the NIS2 directive in
relation to the Critical Entities Resilience (CER)
directive.\footnote{~Directive~(EU) 2022/2557.} This point is important and
somewhat surprising because the NIS2 and CER directives are Siamese twins. In
particular, both directives operate with the same critical entities, the
identification of which should be synchronized in national adaptations to ensure
sound administration and enforcement.

\begin{table*}[th!b]
\centering
\caption{Legal Framings}
\label{tab: legal framings}
\begin{tabularx}{\linewidth}{lXl}
\toprule
Framing towards & Publications \\
\hline
The CER & \cite{springer16, springer26, springer17, springer2, springer23} \\
\cmidrule{2-2}
The CRA & \cite{springer15, springer25, ieee2, springer6, springer14, springer2, elsevier3, springer21} \\
\cmidrule{2-2}
The CSA & \cite{springer20, springer14, elsevier3} \\ 
\cmidrule{2-2}
The CSOA & \cite{springer6, springer2} \\ 
\cmidrule{2-2}
The DSA & \cite{springer21} \\
\cmidrule{2-2}
The DORA & \cite{springer9, springer16} \\
\cmidrule{2-2}
The e-privacy directive & \cite{springer14} \\
\cmidrule{2-2}
The GDPR & \cite{springer6, springer14, springer21} \\
\cmidrule{2-2}
The MDR & \cite{taylorfrancis1} \\
\cmidrule{2-2}
The PSD2 & \cite{springer24} \\
\cmidrule{2-2}
The NIS directive & \cite{springer8, elsevier1, ieee4} \\
\cmidrule{2-2}
The RED & \cite{taylorfrancis1} \\
\cmidrule{2-2}
National laws in the EU & \cite{springer27, springer8, springer1, ieee3} \\
\cmidrule{2-2}
National laws in other countries & \cite{springer27, springer13, ieee4, springer21} \\
\bottomrule
\end{tabularx}
\end{table*}

The fourth point is that NIS2 has perhaps a little surprisingly been frequently
discussed in conjunction with the Cyber Resilience Act
(CRA).\footnote{~Regulation (EU) 2024/2847.} A possible explanation might be
that these two EU laws have been perceived as particularly relevant and
important in terms of cyber security---a sentiment with which the author of this
paper would also agree. Though, with respect to legal background, the two are
rather different because the CRA's background and logic are strongly motivated
by and built upon the EU's product safety laws and consumer protection
jurisprudence in general~\cite{springer15}. Another foundational divergence is
related: the NIS2 directive is on the side of critical infrastructure
protection, including with respect to services vital to the functioning of
European societies, whereas the CRA seeks to improve cyber security of software
and hardware products. Given these fundamental divergences, the rationale behind
many studies has been comparative; to evaluate whether and where there are still
similarities. In this regard, the reporting and communication obligations
imposed by the two laws have been investigated with an argument that
synchronization would again be desirable to the extent
possible~\cite{springer15}. In addition, common traits have been found in terms
of risk analysis and risk management, definitions regarding criticality,
incident handling, and market surveillance provisions~\cite{springer25, ieee2,
  elsevier3}. These research results and arguments thereto are related to the
earlier point about complexity and fragmentation. If the arguments are true,
better coordination during policy-making could have perhaps prevented regulatory
fragmentation, duplication, and general policy incoherence.

Also many other EU laws have been frequently mentioned when discussing
NIS2. Among these is the Cyber Security Act (CSA), the law proposal for a Cyber
Solidarity Act (CSOA), and the so-called e-privacy
directive.\footnote{~Regulation (EU) 2019/881, COM(2023) 209 final, and
Directive 2002/58/EC, respectively.} These further laws lead to the fifth point:
many of the additional laws were often discussed because they were relevant in a
particular context a publication operated. For instance, a publication
investigating the so-called metaverse expectedly discussed also the General Data
Protection Regulation (GDPR), further raising points regarding the Digital
Services Act (DSA).\footnote{~Regulation (EU) 2016/679 and Regulation (EU)
2022/2065, respectively.} Another example would be the several publications
operating in the Internet of things (IoT) domain. Here, not only is the CRA
relevant, but also the older Radio Equipment Directive (RED)
applies.\footnote{~Regulation (EU) 2024/2847 and Directive 2014/53/EU,
respectively.} A similar point applies to medical devices, which are excluded
from the CRA's scope, but to which the Medical Devices Regulation (MDR)
applies.\footnote{~Regulation (EU) 2017/745.} A further example would be the
couple of publications operating in the domain of banking and finance. In this
domain the PSD2 and DORA laws are relevant.\footnote{~Directive (EU) 2015/2366
and Regulation (EU) 2022/2554, respectively.} Finally, there is the NIS2's
predecessor, the old NIS directive.\footnote{~Directive (EU) 2016/1148.} In this
regard, three publications have taken an always sensible approach to investigate
what is really new.

As if the multitude of EU laws would not be enough, particularly legal
scholarship but also others have further investigated national laws in Europe,
including the transpositions of the EU laws, as well as related cyber security
laws in other, non-European countries. The examples include comparisons of cyber
security laws and cyber security governance in Germany and the United
States~\cite{springer27}, Germany and Italy~\cite{springer1}, Canada and the
EU~\cite{springer13}, Australia, Canada, the EU, and the United
States~\cite{ieee4}, and the EU and the United Kingdom~\cite{springer21}. In
addition, there are a couple of publications investigating similar topics with
national case studies focusing on Italy~\cite{springer8} and the Czech
Republic~\cite{ieee3}. Despite some progress, a critical but reasonable
argument can be raised that the existing knowledge is very limited and fragile
regarding particularly the international scene. It also seems unlikely that SLRs
alone could address this limitation. Instead, it might be more reasonable to
tackle the knowledge gap with longer term pursuits involving international
conferences, edited collections, special issues in journals, and by other
related means.

\subsection{Contexts}

The NIS2 directive has been discussed in a number of different contexts, some of
which are directly related to distinct technologies and some others of which are
more generally about cyber security. The study contexts are summarized in
Table~\ref{tab: contexts}. For unpacking the table, it can be started by noting
the couple of publications that have addressed the NIS2 directive in a context
involving industrial control systems, including supervisory control and data
acquisition (SCADA) systems, and related technologies. The more specific
contexts are threat modeling and risk analysis for industrial control
networks~\cite{springer14} and a cyber security analysis of water
towers~\cite{springer10}. Industrial control and SCADA systems are generally
good examples because they are often seen as being the primary technological
components behind critical infrastructures~\cite{Lehto22}. Furthermore, water
towers are a relevant example because both drinking water and waste water are
specified as critical entities in the NIS2 directive. In addition, the energy
sector too is specified as a critical entity; in fact, NIS2 specifies anything
and everything related to electricity, heating and cooling, oil, gas, and
hydrogen as being critical. To this end, the unpacking can be continued by
noting a couple of publications operating in the energy sector context. The
first of these two publications is about using digital twins to monitor smart
grids~\cite{acm1}. The second is a more general policy analysis, including with
respect to NIS2, which recommends policy markers to consider new technologies,
such as artificial intelligence (AI), cyber security training and education, and
improved coordination to tackle cyber threats affecting the energy
sector~\cite{springer4}. There are also other contexts explicitly related to
the Europe's critical infrastructures.

\begin{table*}[th!b]
\centering
\caption{Study Contexts}
\label{tab: contexts}
\begin{tabularx}{\linewidth}{lX}
\toprule
Domain & Publications \\
\hline
Assurance & \cite{springer20} \\
\cmidrule{2-2}
Cloud computing & \cite{springer5} \\
\cmidrule{2-2}
Deception and camouflage & \cite{springer23} \\
\cmidrule{2-2}
Digital twins & \cite{acm1} \\
\cmidrule{2-2}
Energy and smart grids & \cite{acm1, springer4} \\
\cmidrule{2-2}
Industrial control systems & \cite{springer10, springer14} \\
\cmidrule{2-2}
Information sharing infrastructures & \cite{springer19, springer17, acm2} \\
\cmidrule{2-2}
Internet of things & \cite{taylorfrancis1, acm1, springer7, ieee2, elsevier3} \\
\cmidrule{2-2}
Metaverse & \cite{springer21} \\
\cmidrule{2-2}
Security management infrastructures & \cite{ieee5} \\
\cmidrule{2-2}
Smart cities & \cite{springer3} \\
\cmidrule{2-2}
Security awareness & \cite{springer22} \\
\cmidrule{2-2}
Security standards & \cite{springer6, springer14} \\
\cmidrule{2-2}
Telecommunications & \cite{ieee1, elsevier2, springer11, ieee6} \\
\cmidrule{2-2}
Zero trust & \cite{springer9} \\
\bottomrule
\end{tabularx}
\end{table*}

As could be expected, also telecommunications and computer networking have been
a context for discussing the NIS2 directive. These belong to the NIS2's critical
digital infrastructure category, which covers not only electronic
telecommunications networks but also many related critical infrastructures,
including content delivery networks (CDNs), data centers, top-level domain (TLD)
name registries, and domain name system (DNS) service providers. Regarding the
publications operating in this context, local and regional
networks~\cite{ieee6}, cellular 5G networks~\cite{springer11}, and satellite
communication networks~\cite{elsevier2} have been contextualized with respect
to NIS2. Satellite communication networks are an interesting case because these
fall partially to the NIS2's category of critical space technologies. It can be
also mentioned that AI has again been a motivating technology in this context
too; it is seen to enhance monitoring, situational awareness, vulnerability and
anomaly detection, and even so-called self-healing~\cite{ieee6}. As a broad and
arguably somewhat vague concept~\cite{Franke14}, situational awareness is also
a suitable overall theme for characterizing a mixture of distinct publications
dealing with information sharing and associated infrastructures.

The information sharing and situational awareness context involves ambitious
projects trying to combine several layers of European critical infrastructures,
including everything from industrial control and SCADA systems to video
surveillance, drones, and digital twins to improve cross-country and
cross-sectoral risk analysis and incident management~\cite{springer19}. In a
similar vein, there is also a large-scale project trying to address the risk
analysis obligations imposed by the CER and NIS2 directives upon critical
infrastructure operators by means of collecting data from legacy systems and new
sensors, improving communication infrastructures between stakeholders, and
providing emergency response plans~\cite{springer17}. A further project seeks
to develop a pan-European incident sharing platform~\cite{acm2}. All of these
three projects have been funded through the EU's Horizon Europe programme.

In a somewhat similar context, it is worth mentioning a publication that
designed a multi-layered security management platform for managing cloud
computing infrastructures, edge computing, and IoT
devices~\cite{ieee5}. Another comparable design envisioned combining data from
SCADA systems, security information and event management (SIEM) systems, and IoT
devices, again pitching AI as the silver bullet~\cite{ieee2}. A further
publication designed digital twins with the help of IoT
devices~\cite{acm1}. These designs are good examples because they demonstrate
how the IoT context might be seen as being related to critical
\text{infrastructures---although} IoT devices do not obviously belong to the
NIS2's categorization of critical entities, cloud computing is defined as a
critical entity. A similar reasoning has been behind comprehensive legal
analyses about the EU's new regulatory impacts upon IoT
devices~\cite{springer7, taylorfrancis1}. Despite these legal analyses and
technical designs, it still remains generally unclear to which extent---if at
all---the NIS2 directive applies in the IoT context. To this end, some
publications have seen the CRA as a more relevant regulation in this
context~\cite{elsevier3}. An analogous point has been raised with respect to
metaverse, virtual reality, and augmented reality
technologies~\cite{springer21}. All in all, these issues demonstrate
difficulties in defining, demarcating, and conceptualizing critical
infrastructures through technologies.

The difficulties can be related to the so-called interconnectedness problem in
the critical infrastructure literature; many critical infrastructures depend on
each other, opening a risk of cascading failures, among other
things~\cite{Harasta18, Hurst24, Little02}. The interconnectedness problem is
seen also in a publication addressing smart cities, which were framed by
connecting these to technological infrastructures and the transport, healthcare,
public administration, and many other sectors~\cite{springer3}. The problem
becomes even more convoluted once software is taken into account; not only are
SCADA or situational awareness systems powered by software but practically all
the critical entities listed in the NIS2 directive are dependent on software,
and software too typically contains dependencies~\cite{Ruohonen24JSS}. The
difficulties can be also used to backup an argument that the EU's conception of
critical infrastructure is not entirely sectoral any more because cloud
computing, the DNS, CDNs, and many related technologies are defined as critical
entities in the NIS2 and CER~directives. A similar argument has been raised in a
publication addressing cloud computing in the NIS2 context. Although the NIS2
directive's category of critical digital infrastructure was motivated by the
policy-makers with not only its overall criticality but also with its
dependencies, no details are available regarding how assessments and
measurements were done~\cite{springer5}. This criticism aligns with a further
critical argument that the policy-makers might have defined too many sectors as
critical~\cite{springer2}. Having said that, the SLR sample contains also
several publications that are not explicitly related to any specific technology
or sector. These publications deserve a brief elaboration too.

Related to the legal framings discussed in the previous section are further
framings toward technical standards. In this regard, it can be noted that the
NIS2's recitals 58 and 79 motivate incident management, vulnerability
coordination, threat analysis, and general cyber security measures with explicit
mentions of the international ISO/IEC 30111, ISO/IEC 29147, and ISO/IEC 27000
series of standards. The new European NIS2 cooperation group is also tasked to
exchange best practices and information related to standards according to
Article~14(4)(c). Furthermore, Article~25 encourages the member states to comply
with European and international standards when implementing cyber security risk
management measures specified in Article 21. However, it can be noted that the
NIS2 directive does not explicitly entail the development of specific European
standards, which are an essential part of the CRA and conformance with it. With
these notes, it is understandable that a couple of publications have discussed
NIS2 in relation to standards. Among the standards discussed are the noted
ISO/IEC standards~\cite{springer6}. In addition, the ISA/IEC 62443 standard and
standards developed by the National Institute of Standards and Technology (NIST)
in the United States have been discussed and used~\cite{springer14}. These
publications align with a more general take on assurance particularly with
respect to software~\cite{springer20}. Standards and assurance are important
topics for further evaluation work once the NIS2 directive has been fully
implemented.

Finally, the SLR protocol captured three publications that do not fit well into
the previous thematic categorizations. The first publication is about designing
the so-called zero trust security model in a context involving national CSIRTs,
security operations centers (SOCs), and data breaches~\cite{springer9}. In this
regard, it is worth recalling that CSIRTs are the essential public sector
organizations in the NIS2 directive, the CRA, and related EU laws, whereas SOCs
are related to the CSOA law proposal. The second publication is about cyber
security awareness, which is seen to be linked to the noted situational
awareness technologies in the NIS2 context~\cite{springer22}. The third
publication is more about physical security; it discusses camouflaging critical
infrastructures in order to prevent threats related to theft, sabotage,
terrorism, and related human-driven endeavours~\cite{springer23}. Such threats
and risks are a good way to end the review because they underline that cyber
security is only a part of the overall security conundrum in today's Europe.

\section{Conclusion}\label{sec: conclusion}

The paper presented a systematic literature review on the new NIS2
directive. The directive is an important part in the EU's regulatory cyber
security framework. It addresses particularly the cyber security of critical
infrastructures in Europe. According to the review presented, the increased
complexity and regulatory fragmentation are visible also in the literature in a
sense that many publications have framed and discussed the NIS2 directive in
relation to numerous other EU laws. It seems reasonable to conclude that the
overall judicial crux will motivate further work for years to come.

The publications reviewed have also discussed the NIS2 directive in various
different contexts. Among these are traditional critical infrastructure
contexts, such as those related to the energy and water sectors, industrial
control and SCADA systems, and telecommunications. Many publications have also
sought to design and develop new infrastructures and platforms for better
information sharing and situational awareness. In addition, various
technologies, whether IoT devices, AI, smart cities, or digital twins, have been
discussed in relation to NIS2 alongside standards and miscellaneous cyber
security topics. Given the eleven critical sectors considered in the NIS2 and
CER directives, it seems again safe to assume that these sectors will motivate a
lot of further work concentrating on particular contexts. For instance, to put
aside an odd publication dealing with satellite communications, none of the
publications reviewed discussed space technologies, which are certainly an
interesting contextual area to study also from a cyber security perspective in
the future. Nor was there a publication in the sample that would have focused on
public administration as a critical~entity.

In addition to these concluding remarks, six points can be raised about research
gaps and topical areas that would require or benefit from further work. These
six points should not be taken to imply that there would not be more; the NIS2
directive is a comprehensive law that likely opens also other further research
possibilities.

First, it is worth revisiting the argument about the NIS and CER directives
having expanded the scope of critical infrastructures. Both legal and more
theoretical further research is warranted in this regard. A generic problem with
expansions is well-recognized in the academic literature; terms such as
conceptual expansion~\cite{Spinuzzi11} and conceptual
stretching~\cite{Ruohonen21MIND} have been used to elaborate the dangers
involved with enlarged definitions; once almost everything belongs to a
definition or a theoretical concept, the definition or concept starts to lose
its qualifying characteristics. In other words: if almost everything is seen as
critical, then not much is available to demarcate non-critical entities from
critical entities. Here, the problem's kernel can be seen to originate from the
concept of criticality, and the EU and its regulations are far from being the
only ones facing the problem~\cite{Ruohonen24JSS}. As was discussed, further
theoretically oriented work is required particularly regarding the EU's
traditional sectoral definition for critical infrastructures and its relation to
the new digital infrastructure category present in the CER and NIS2 directives.

Second, it seems sensible to argue that risk management would deserve further
contributions too in the NIS2 context. Although there is plenty of existing work
in this area with respect to standards and technical solutions, the NIS2's whole
scope has not been addressed. In particular, the directive's Article~21(2)
mandates ``an all-hazards approach'' covering ten distinct risk management
areas. These include traditional areas, such as risk analysis, incident
handling, vulnerability coordination and disclosure, but also supply chain
security, so-called cyber hygiene practices, authentication, encryption, human
resources security, and even business continuity should be covered. Of these
areas, supply chain risk management is particularly interesting as it is
directly related to the interconnectedness problem and cascading
failures~\cite{Sanders23}. The all-hazards approach envisioned would also
benefit from further work investigating and modeling how risks in the different
areas are potentially connected to each other.

Third, none of the studies captured by the SLR protocol addressed or even
discussed the databases and registries established and mandated by the NIS2
directive. Although only time will tell how these will function and what impact
they will have, these can be seen as an important part of the directive in many
ways. Thus, to begin with, the NIS2's Article~12(2) introduces a new European
vulnerability database. Once established and functional, this database requires
evaluations and assessments already because of the well-known problems that have
plagued vulnerability databases in general~\cite{Anwar22, Massacci13}. There
are also opportunities for further work regarding potential international
cooperation, interoperability between vulnerability databases, new vulnerability
concepts brought by the CRA, and the effects of these upon vulnerability
disclosure, coordination, and mitigation~\cite{Ruohonen24IFIPSEC}. Then, to
proceed, Article~27 in the NIS2 directive mandates registrations from all
entities belonging to the digital infrastructure category to a centralized
database managed at the EU-level. In addition to basic information, such as
contact details, the entities must supply their Internet protocol address
ranges. Furthermore, the subsequent Article~28 in the directive seeks to improve
the security, stability, and resilience of the DNS by mandating the member
states to collect registration data from not only TLD name registries but also
from all conventional domain name registrations. Given the importance of domain
name registration data in analyzing cyber attacks and resolving cyber
crimes~\cite{Maroofi20, Shi18}, this new obligation can be seen as particularly
relevant in the NIS2's database~context.

Fourth, many of the publications reviewed operated in a context involving new
technologies. In this context, it is worth recalling the publication that
designed data collection from legacy systems. The reason is that many critical
infrastructures either are legacy systems themselves or depend on legacy
systems, which may increase cyber security risks and typically constrain the
adoption of new technologies, including AI solutions~\cite{Hurst24,
  Maglaras18}. This point should not be taken as an explicit criticism about the
publications reviewed themselves. Instead, it is more about the NIS2 directive
itself, which, in its recitals 51 and 89, encourages the member states to adopt
innovative technologies, including AI, to tackle cyber security threats,
including with respect to critical infrastructure protection. Given such an
encouragement, further research is required about the presumably difficult
interplay between critical infrastructures, legacy systems, and
new technologies.

Fifth, many publications reviewed also designed and envisioned large-scale
information sharing and situational awareness infrastructures. In addition to
the previous point, it is worth remarking the potential practical obstacles
constraining the adoption and deployment of such infrastructures. There is a
good reason for this remark: despite some progress, technical obstacles,
including those related to standardization and interoperability, trust, and
other related factors have constrained the sharing of cyber security information
throughout the world~\cite{springer2, Skopik16, Zibak19}. Further research is
required about such obstacles potentially affecting the NIS2's implementation
particularly with respect to union-wide, pan-European information sharing and
associated critical infrastructure protection.

Last, it is worth returning to the evaluation research already noted. While it
is impossible to envision which particular areas will be especially relevant for
evaluations, administration and enforcement are always relevant in the context
of regulations. To this end, it seems again sensible to argue that the increased
legal complexity and fragmentation will be felt also among public
administrations, whether national or European. Administrative efficiency would
thus likely offer a plausible research topic, whether examined by the means of
surveys or interviews. Analogously to the GDPR's administration and
enforcement~\cite{Ruohonen22IS}, an alternative path would involve examining
future administrative fines and other penalties from non-compliance. Here, it is
worth mentioning that both the NIS2 directive and the CRA regulation entail
potentially severe financial consequences from non-compliance. While fines,
penalties, and non-compliance in general are essentially a negative way to
approach regulations---nothing is said about compliance and good practices in
general, they might still in the future reveal some particular bottlenecks
potentially affecting the private sector operators of the Europe's critical
infrastructures. With these points in mind, further work, perhaps in a form of
university-industry collaborations, is required on the means by which compliance
can be reasonably gained and proved.

\bibliographystyle{splncs03}

\end{document}